\documentclass{iaus}
\usepackage{graphicx}

\title[Chromospheric activities] 
{Chromospheric variabilities of M active stars based on Guoshoujing Telescope
}

\author[L.Y. Zhang, et al.,]   
{ Q. F. Pi$^1$, L. Y. Zhang$^1$, J. R. Shi$^2$, H. Wu$^2$, 
 Y. H. Zhao$^2$, A. L. Luo$^2$, J. K. Zhao$^2$, A. Y. Zhou$^2$, X. S. Fang$^1$, \and Lamost Collaboration$^2$}

\affiliation{$^1$ College of Science / Department of
Physics, Guizhou University, Guiyang 550025, China\\ email: {\tt Liy\_zhang@hotmail.com} \\[\affilskip]
$^2$ National Astronomical Observatories, Chinese Academy of Sciences, Beijing 100012, China  }

\pubyear{2014}
\volume{298}  
\pagerange{XX -- YY}
\jname{IAUS\,298 Setting the scence for Gaia and LAMOST}
\editors{S. Feltzing, G. Zhao, N.\,A. Walton \& P.\,A. Whitelock, eds.}
\begin{document}

\maketitle

\begin{abstract}
 For the M active catalogue of Guoshoujing Telescope (LAMOST), 933 sources are presented in at least two exposures. We found that many M active stars show chromospheric variabilities in the Ca H, H$_{\alpha}$, H$_{\beta}$, and H$_{\gamma}$ lines on short or long timescales.

\keywords{stars: M, stars: activity, stars: chromospheres, stars: spectra}
\end{abstract}


{Chromospheric variabilities of M dwarf stars are very interesting (Hilton et al. 2012; Kruse et al. 2012; etc). We found 6391 M active candidates (Zhang et al. 2012, 2013) from LAMOST spectral survey (Cui et al. 2012; Zhao et al. 2012; Luo et al. 2012; Deng et al. 2012). 933 sources are found in at least two exposures, in which 193 with more than 3 three exposures. These spectra obtained at different times can be used to examine chromospheric variability. We measured their equivalent widths of the Ca H, H$_{\alpha}$, H$_{\beta}$, and H$_{\gamma}$ lines using the program of Hawley et al. (2002) and West et al. (2011). The wavelength regions used for the continuum and lines regions are similar to that of Hilton et al. (2012). We found that chromospheric activities of most active stars had changed on either short or long time scales. Figure 1 shows an example of LAMOST J045556.06+303620.6 in Ca II H, H$_{\alpha}$, H$_{\beta}$, and H$_{\gamma}$ lines.

\vspace{-0.5cm}
 \begin{figure}[h]
\center
\includegraphics[width=14.50cm,height=3.2cm]{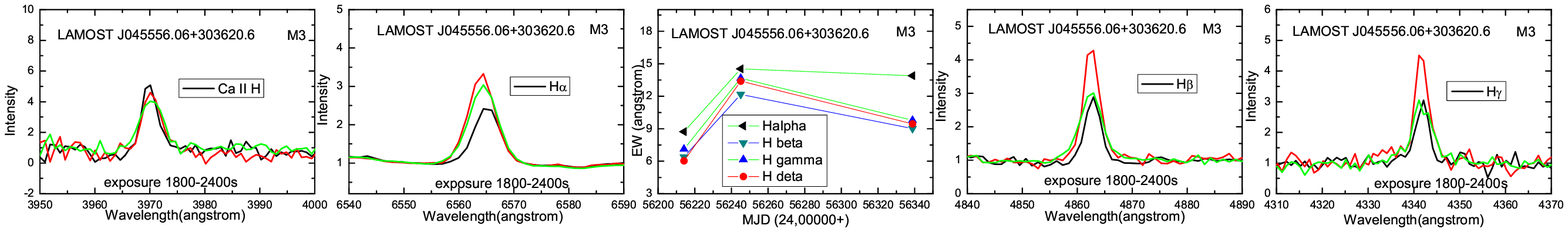}
\caption{The spectra variation and EWs light curves of LAMOST J045556.06+303620.6.}
\end{figure}
\vspace{-0.1cm}
{\bf Acknowledgements}
This work is supported by the NSFC of No 10978010 and 11263001.

%
%
%
%

\end{document}